\documentclass[aps,pra,reprint,superscriptaddress]{revtex4-1}

\usepackage{hyperref}
\hypersetup{
	pdfborder={0 0 0},
	pdfnewwindow=true,      
	colorlinks=true,       
	linkcolor=blue,          
	citecolor=blue,        
	filecolor=blue,      
	urlcolor=blue,          
}
\usepackage{graphicx} 
\usepackage{amsmath} 
\usepackage{amsfonts} 
\usepackage{xcolor}
\usepackage[normalem]{ulem} 

\usepackage{pdfpages} 
\makeatletter
\AtBeginDocument{\let\LS@rot\@undefined}
\makeatother
\usepackage{pgffor}

\newcommand{\Eq}{Eq.~}

\newcommand{\Fig}{Fig.~}

\renewcommand{\vec}[1]{\boldsymbol{#1}}
\DeclareMathAlphabet{\mathsfsl}{OT1}{cmss}{m}{sl}

\newcommand{\mi}{\mathrm{i}}
\newcommand{\dif}{\mathrm{d}}

\newcommand{\llangle}{\left\langle\!\left\langle}
\newcommand{\rrangle}{\right\rangle\!\right\rangle}

\begin{document}


\title{Geometric effects in the effective-mass theory \\and topological optical superlattices}


\author{Chao-Kai Li}
\affiliation{International Center for Quantum Materials, School of Physics, Peking
	University, Beijing 100871, China}

\author{Qian Niu}
\affiliation{Department of Physics, The University of Texas, Austin, Texas 78712-0264, USA}

\author{Ji Feng}
\email{jfeng11@pku.edu.cn}
\affiliation{International Center for Quantum Materials, School of Physics, Peking
	University, Beijing 100871, China}
\affiliation{Collaborative Innovation Center of Quantum Matter, Beijing 100871,
	China}
\affiliation{CAS Center for Excellence in Topological Quantum Computation, University of Chinese Academy of Sciences, Beijing 100190, China}


\date{\today}

\begin{abstract}
Cold atoms tailored by an optical lattice have become a fascinating arena for simulating quantum physics. In this area, one important and challenging problem is creating effective spin-orbit coupling (SOC), especially for fashioning a cold atomic gas into a topological phase, for which prevailing approaches mainly rely on the Raman coupling between the atomic internal states and a laser field. Herein, a strategy for realizing effective SOC is proposed by exploiting the geometric effects in the effective-mass theory, without resorting to internal atomic states. It is shown  that the geometry of Bloch states can have nontrivial effects on the wave-mechanical states under external fields, leading to effective SOC and an effective Darwin term, which have been neglected in the standard effective-mass approximation. It is demonstrated that these relativisticlike effects can be employed to introduce effective SOC in a two-dimensional optical superlattice, and induce a nontrivial topological phase.
\end{abstract}


\maketitle


In recent years, topological insulators, in which spin-orbit coupling (SOC) plays a central role, have become a focal spot in condensed matter-physics~\cite{Hasan2010,Qi2011}. Cold-atom systems, due to their extreme cleanness and high controllability, are expected to be an ideal platform for exploring various phenomena related to topological matter as well as the interplay between topology and  interaction.  In order to synthesize a topological insulator in cold-atom systems, and to explore the many intriguing effects of SOC itself on cold atoms that have no analog in solid-state systems, it is important to create artificial SOC in neutral atoms~{\cite{Dalibard2011,Zhai2012,Galitski2013,Goldman2014,Zhai2015}}. This is not an easy task, although quite a few schemes have been theoretically proposed~\cite{Dudarev2004,Osterloh2005,Ruseckas2005,Liu2009,Juzeliunas2010,Zhang2010,Campbell2011,Anderson2012,Xu2013,Anderson2013,Liu2014}. Based on these schemes, effective SOC has been achieved experimentally in both one-dimensional~\cite{Lin2011,Cheuk2012,Wang2012,Zhang2012,Williams2013,Qu2013,Olson2014} and two-dimensional~\cite{Huang2016,Wu2016} cold-atom systems. These methods mainly rely on the coupling of the atom's internal states through an additional Raman laser field. Because different types of atoms have different internal excitation structures, these methods are limited by the availability of suitable internal levels that afford Raman coupling and by the lifetime of the excited states. Another problem is that the spontaneous emission between internal states leads to the heating of quantum gases or atom loss.

In this Rapid Communication, an approach is proposed for the realization of effective spin-orbit coupling by exploiting the geometric effects in the effective-mass theory. The geometric effect is revealed in relativisticlike terms in a new effective-mass equation derived herein. When applied to a cold-atom system in an optical superlattice, this theory leads to effective relativistic effects without  utilizing the internal structure of the atoms, because the SOC effect emerges from the geometry of the Bloch bands of the host lattice. Consequently, this approach is expected to be versatile for various types of atoms, free of heating and atom loss due to spontaneous emission. A model 2D optical superlattice is devised to demonstrate the effective SOC effect, which is shown to indeed lead to a topologically nontrivial phase. This approach can be naturally applied to 3D cases.

The effective-mass theory~\cite{Slater1949,Adams1952,Kittel1954,Koster1954,Luttinger1955}, initially formulated for an elementary understanding of  shallow impurity states in semiconductors, has played a vital role in semiconductor physics. It also has found wide applications in the quantum physics associated with general imperfections of a lattice such as intercalation, quantum dot, and boundaries , and with the motion of electrons in a crystal under external fields~\cite{DiVincenzo1984, Li1996, Bastard1981, Bastard1982, Burt1992, Matsumura1998,Kohn1959,Wannier1960}.
However, the geometric effects of Bloch electrons in association with the Berry curvature and quantum metric, which have drawn a great deal of attention for decades, are conspicuously absent in the traditional effective-mass equation.
Consequently, we start by extending the effective-mass approximation to a higher order aiming to capture the geometric effects arising from the Bloch bands. It is shown that the geometric effects are embodied in the Berry curvature and quantum metric tensor, which lead to unusual corrections to the traditional effective-mass approximation, bearing a remarkable resemblance to the SOC and Darwin term in the nonrelativistic approximation of the Dirac equation~\cite{Sakurai}. Though the effective-mass equation has traditionally been developed in an electronic structure context, as our cold-atom application demonstrates, the higher-order effective-mass equation can be profitably employed to understand and tailor wave-mechanical states associated with an imperfection on a host lattice, including crystals, optical lattices, and artificial superstructures.

In deriving the standard effective-mass equation, a key quantity is the inner product of the cell-periodic parts of Bloch wave functions at different $\vec k$ points, $ \langle u_{n\vec{k}} | u_{n'\vec{k}'}\rangle $. This overlap reflects the metric change of the Hilbert spaces of distinct $\vec k$ points when the Brillouin zone is viewed as a manifold, and is associated with such geometric quantities as the Berry curvature and quantum metric~\cite{Berry1984, Provost1980,Berry89,Xiao10}. Nonetheless, this important geometric quantity is taken as $ \delta_{nn'} $ in the traditional effective-mass approximation~\cite{Luttinger1955}; that is, at this level of approximation a trivial geometry is always adopted, which can otherwise have a highly nontrivial manifestation. For example, the geometric effect has been suggested to play an important role in the excitonic states in 2D semiconductors~\cite{Zhou2015,Srivastava2015}.

We begin by formulating the impurity problem. For a perfect lattice with Hamiltonian $H_{0}$ respecting discrete translational symmetry, its eigenstates
are the Bloch wave functions $|\psi_{n\vec{k}}\rangle$, with corresponding
eigenenergies $\varepsilon_{n}(\vec{k})$, $n$ being the energy band index
and $\hbar\vec{k}$ the crystal momentum. When an impurity is placed in
a perfect host lattice,  the stationary Schr\"odinger equation is
\begin{equation}
	\left(H_{0}+U\right)|\Psi\rangle=E|\Psi\rangle ,
	\label{eq:scheq}
\end{equation}
where the impurity potential $U$ breaks the discrete translational symmetry of the host lattice.
The eigenstate
$|\Psi\rangle$ can be expanded as
\begin{equation}
	|\Psi\rangle=\sum_{n}\int_{\text{BZ}} [\dif\vec{k}]F_{n\vec{k}}|\psi_{n\vec{k}}\rangle \text{,}
	\label{eq:expand by Bloch wave functions}
\end{equation}
in which $F_{n\vec{k}}$ is the expansion coefficient, and we use the notation 
$[\dif \vec{k}] \rightarrow  V_{\text c} \dif ^d \vec{k} /(2\pi)^d$ 
with $V_c$ being the volume of a primitive cell and $d$
the dimensionality. The expansion
(\ref{eq:expand by Bloch wave functions}) leads to a set of linear equations of $F_{n\vec{k}}$,
\begin{equation}
	\varepsilon_{n}(\vec{k})F_{n\vec{k}}+\sum_{n'}\int_{\text{BZ}} [\dif \vec{k} ]\left\langle \psi_{n\vec{k}}\left|U\right|\psi_{n'\vec{k}'}\right\rangle F_{n'\vec{k}'}
	= E † F_{n\vec{k}}.
	\label{eq:Schrodinger eqn of Fnk}
\end{equation}

Up to this point, the formulation is exact given the Schr\"odinger equation (\ref{eq:scheq}). Several standard approximations are introduced in the traditional effective-mass approximation~\cite{Adams1952,Luttinger1955}.  The impurity potential $U(\vec{r})$ is assumed to be slowly varying in space, and correspondingly,  $F_{n\vec{k}}$ is concentrated near an energy-band extremum (or a valley) at $ \vec{k}_0 $. It follows that the umklapp scattering can be neglected. 
With these assumptions, we have
\begin{equation}
	\label{eq: matrix element of U}
	\left\langle \psi_{n\vec{k}}\left|U\right|\psi_{n'\vec{k}'}\right\rangle \approx\frac{\left(2\pi\right)^{d/2}}{V_{c}}U\left(\vec{k}-\vec{k}'\right)\left\langle u_{n\vec{k}}|u_{n'\vec{k}'}\right\rangle _{\text{cell}},
\end{equation}
where $|u_{n\vec{k}}\rangle$ is the cell-periodic part of $|\psi_{n\vec{k}}\rangle$. The subscript ``cell'' emphasizes that the integral is to be taken over a primitive cell.

Consequently, only the matrix elements $ \langle \psi_{n\vec{k}} | U | \psi_{n'\vec{k}'} \rangle $ for $ \vec{k} \approx \vec{k}' \approx \vec{k}_{0} $ make a significant contribution in \Eq (\ref{eq:Schrodinger eqn of Fnk}). To the lowest order, $ \langle u_{n\vec{k}} | u_{n'\vec{k}'} \rangle \approx \langle u_{n\vec{k}_{0}} | u_{n'\vec{k}_{0}} \rangle = \delta_{nn'} $ in \Eq (\ref{eq: matrix element of U}), as has been adopted in the traditional effective-mass  approximation~\cite{Adams1952,Luttinger1955}. However, since the quantity $\langle u_{n\vec{k}} | u_{n'\vec{k}'} \rangle$ contains the geometric relation between $\vec{k}$ and $\vec{k}'$, this approximation amounts to endowing the Brillouin zone, here viewed as a manifold, with a completely trivial geometry.

Now we expand $|u_{n\vec{k}}\rangle$ to second order in $ \delta \vec{k} = \vec{k} - \vec{k}_{0} $.
It is crucial to keep the expansion up to second order to retain the geometric effects. As detailed in the Supplemental Material (SM)~\cite{SM}, with a suitable gauge choice in the one-band scenario, a new effective-mass equation can be obtained,
\begin{equation}
	\begin{split}
		& \Big[\varepsilon_{n}\left(-\mi\vec{\nabla}\right)_{\vec{k}_{0}}+U\left(\vec{r}\right)-\frac{1}{2}\vec{\Omega}_{n}\left(\vec{k}_{0}\right)\cdot\vec{\nabla} U\times\left(-\mi\vec{\nabla}\right) \\
		& +\frac{1}{2}g_{n,\alpha\beta}\left(\vec{k}_{0}\right) \left(\partial^{\alpha}\partial^{\beta}U \left(\vec{r}\right)\right)\Big]F_{n}\left(\vec{r}\right) =EF_{n}\left(\vec{r}\right) \text{,}
		\label{eq:corrected effective mass equation}
	\end{split}
\end{equation}
in which
\begin{equation}
	\vec{\Omega}_{n}\left(\vec{k}_{0}\right)=\mi\left\langle \vec{\nabla}_{\vec{k}}u_{n}\left|\times\right|\vec{\nabla}_{\vec{k}}u_{n}\right\rangle _{\vec{k}_{0}}
\end{equation}
is the Berry curvature of the Bloch band~\cite{Berry1984,Xiao10},
\begin{equation}
	g_{n,\alpha\beta}\left(\vec{k}_{0}\right)=\frac{1}{2}\left(\left\langle \partial_{k_{\alpha}}u_{n}\left|Q\right|\partial_{k_{\beta}}u_{n}\right\rangle + \alpha \leftrightarrow \beta \right)_{\vec{k}_{0}}
\end{equation}
is the quantum metric~\cite{Provost1980}, and $Q=1-|u_{n}\rangle\langle u_{n}|$ is a projection operator. These quantities reflect the geometric effects of the Bloch bands and lead to two additional terms to the traditional effective-mass equation, with which the new effective-mass equation has a striking resemblance to the nonrelativistic approximation of the Dirac equation~\cite{Sakurai}. Remarkably, these relativisticlike terms correspond respectively to the spin-orbit coupling and Darwin term. The former can be procured heuristically by requantization of the effective semiclassical dynamics, but the latter cannot~\cite{Chang2008}. The present approach has consistently stayed quantum mechanical and is systematic and therefore extensible, and the effective Hamiltonian here involves only gauge-invariant geometric quantities, i.e. the Berry curvature and quantum metric.

The physical meaning of the effective spin-orbit coupling (ESOC) in the new effective-mass equation can be made clearer in a multivalley scenario. Because the impurity potential is assumed
to be slowly varying, the intervalley coupling---which involves a large crystal momentum transfer---can be neglected in
the lowest-order approximation. Hence, each valley has an independent effective-mass equation (\ref{eq:corrected effective mass equation}), which
differs from each other by the effective mass, Berry curvature,
and quantum metric at the valleys. Intriguingly, in the ESOC
term the role of ``spin'' is now played by the valleys, and the strength of the
``spin''-orbit coupling is proportional to the magnitude of the
Berry curvature. The appearance of a quantum metric in the new effective-mass equation reflects in part the anisotropy of the host lattice.

\begin{figure}
	\begin{center}
		\includegraphics[width=4.5cm]{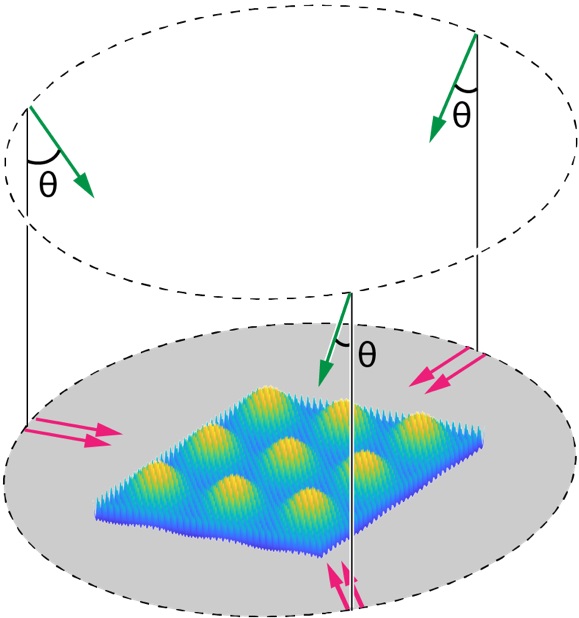}
		\caption{\label{fig: optical_lattice}A proposed experimental setup for the generation of potential $ V(\vec{r}) + U(\vec{r}) $ by an optical lattice. The colored arrows indicate nine laser beams, which can be obtained by three laser sources. Six of them (pink) lie in the same plane, generating the ``host lattice'', and the other three lasers (green) point obliquely from out of the plane, generating the ``impurity'' superlattice. A horizontal trapping potential is used to make the optical lattice two dimensional.}
	\end{center}	
\end{figure}

Having the new effective-mass equation, we now demonstrate that the ESOC can be exploited to create a 2D topological optical superlattice. The host lattice is taken to be a 2D honeycomb lattice with a staggered $ A $-$ B $ sublattice potential created by the interference of six coplanar laser beams, as shown by the six pink arrows in \Fig \ref{fig: optical_lattice}. A horizontal trapping potential is used to confine the atoms to a 2D plane. These six laser beams can be divided into two groups. Each group consists of three laser beams with the same frequency and with a $ 120^{\circ} $ angle with each other, generating a triangular lattice by interference. These six laser beams can be generated by three laser sources by the use of acousto-optic modulators, which detune the frequencies between the two groups of laser beams, so as to avoid interference between the two groups, and to control the relative strength of $ V_{A} $ and $ V_{B} $. Together, the six laser fields generate a host lattice potential
\begin{widetext}
\begin{eqnarray}
	\label{eq: host lattice potential}
	V(x,y)&=&-2V_{A}\left\{ \cos\left[2\pi\left(\frac{x}{a}-\frac{1}{\sqrt{3}}\frac{y}{a}-\frac{1}{3}\right)\right]+\cos\left[2\pi\left(\frac{x}{a}+\frac{1}{\sqrt{3}}\frac{y}{a}+\frac{1}{3}\right)\right]+\cos\left[2\pi\left(\frac{2}{\sqrt{3}}\frac{y}{a}-\frac{1}{3}\right)\right]\right\}  \nonumber\\
	&&-2V_{B}\left\{ \cos\left[2\pi\left(\frac{x}{a}-\frac{1}{\sqrt{3}}\frac{y}{a}+\frac{1}{3}\right)\right]+\cos\left[2\pi\left(\frac{x}{a}+\frac{1}{\sqrt{3}}\frac{y}{a}-\frac{1}{3}\right)\right]+\cos\left[2\pi\left(\frac{2}{\sqrt{3}}\frac{y}{a}+\frac{1}{3}\right)\right]\right\} . \nonumber\\
\end{eqnarray}
\end{widetext}
The two lines of \Eq (\ref{eq: host lattice potential}) are generated by the two groups of laser beams, respectively. The parameter $ a $ is the lattice constant of this host lattice. Note that if $ V_{A}=V_{B} $, the six cosines in $ V(x,y) $ can be combined into three cosines by a sum-to-product identity, and hence three laser beams are enough to generate a honeycomb lattice with inversion symmetry~\cite{Grynberg1993}. The difference of $ V_{A} $ and $ V_{B} $ breaks inversion symmetry, and  results in direct band gaps at the hexagonal Brillouin zone corners $K$ and $K'$, where the magnitudes of Berry curvature are large~\cite{SM}. For what follows, we will focus on the conduction-band edges at $K$ and $K'$, which are referred to as valleys.

We begin by applying \Eq (\ref{eq:corrected effective mass equation}) to analyze the energy levels in the single ``impurity'' limit, for the two valleys $\vec{k}_0 = K,K'$. Consider that a large-scale trapping potential $U\left(\vec{r}\right)$ is imposed on the host lattice playing the role of an ``impurity'' potential. 
Assuming that the trapping potential $U\left(\vec{r}\right)$
in \Eq (\ref{eq:corrected effective mass equation}) has 2D
rotational symmetry,
the solutions can be classified by their angular
momenta $L_{z}=l\hbar$, where $l$ is the azimuthal quantum number. For the 2D honeycomb host lattice, the Berry curvature $ \vec{\Omega} $ has the same magnitude and opposite signs at $ K $ and $ K' $~\cite{SM}, while the quantum metric $ g $ is the same for both valleys. Hence, the ESOC
leads to splitting of the otherwise degenerate states with
$L_{z}=\pm l\hbar$, $ l \ne0$. The lowest-energy states are two ``$ 1s $''
states originating from the two valleys with vanishing angular momenta.  Neither the ESOC nor the effective Darwin term can lift the degeneracy of the two $ 1s $ states. In the absence of intervalley scattering, the two $ 1s $ states of $K$ and $K'$ valleys are degenerate, and we can treat them as  the two spin projections,  $|\uparrow\rangle$
and $|\downarrow\rangle$,  of a ``pseudospin''-1/2.

\begin{figure}[!h]
	\begin{center}
		\includegraphics[width=8.5cm]{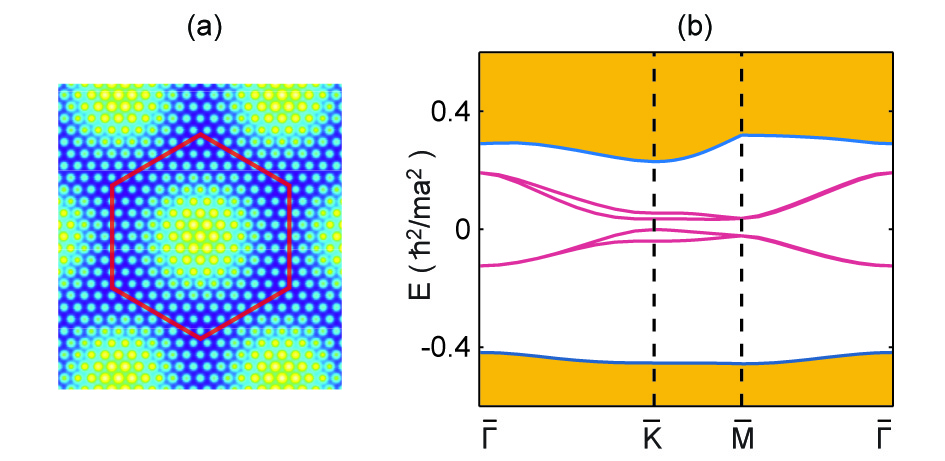}
		\caption{\label{fig: superlattice_bands_sin}(a) Schematic of the optical superlattice with potential $ V(x,y) + U(x,y) $. A supercell is indicated by the red hexagon. The small dark blue points are the positions of the valleys of the host lattice potential $ V(x,y) $, and the small yellow points are the positions of the peaks of $ V(x,y) $; so are the large dark blue and yellow areas to the large-scale potential $ U(x,y) $. (b) Four minibands (red) are formed in the band gap of the host lattice.  The orange regions correspond to the areas occupied by bulk bands of the host lattice.}
	\end{center}
\end{figure}

Then we analyze the case when a large-scale 2D honeycomb superlattice, which can be viewed as a periodic array of the above-mentioned trapping potentials, is superimposed on the host lattice. The total large-scale
potential $ U(\vec{r}) $ [\Fig \ref{fig: superlattice_bands_sin}(a)] is no longer fully rotational symmetric, but reduced to
the symmetry of a honeycomb superlattice. Because of the hoppings among the trapping sites, the ``impurity levels'' become dispersive minibands. With the two $ 1s $ states $|\uparrow\rangle$
and $|\downarrow\rangle$ on each trapping site as the basis, a tight-binding (TB) Hamiltonian for the large-scale honeycomb superlattice can be built, 
\begin{eqnarray}
H &=& \Delta\sum_{i\mu}\alpha_{i}c_{i\mu}^{\dagger}c_{i\mu}-t\sum_{\left\langle ij\right\rangle \mu}c_{i\mu}^{\dagger}c_{j\mu} \nonumber\\
&&+\frac{\mi}{3\sqrt{3}}\sum_{\llangle ij \rrangle \mu\mu'}\left(\lambda+\alpha_{i}\xi\right)\nu_{ij}c_{i\mu}^{\dagger}\left(s_{z}\right)_{\mu\mu'}c_{j\mu'}\label{eq:Kane-Mele, real space} \text{,}
\end{eqnarray}
in which $c_{i\mu}^\dagger$ creates a particle with pseudospin $\mu$ on trapping site $i$, and $\alpha_{i}= +1 /-1$
for trapping site $i$ lies on the $ A $/$ B $ sublattice. $2\Delta$ is the on-site energy difference of the two
sublattice originating from the inversion symmetry breaking of the
host lattice. The hopping matrix element of the nearest neighbors $\langle ij\rangle$ is denoted as $t$, and 
$\left(\lambda\pm\xi\right)$ corresponds to the matrix elements of
the ESOC between next nearest neighbors $\llangle ij \rrangle $ on the $ A $ sublattice
(+) or $ B $ sublattice ($-$), with the difference $2\xi$ originating from
the inversion symmetry breaking. $\nu_{ij}=+1$
if the direction of $j\rightarrow i$ is counterclockwise inside a hexagon of the honeycomb superlattice
and $-1$ if clockwise, and $s_{z}$ is the $ z $ Pauli
matrix for $\uparrow$ and $\downarrow$. The Hamiltonian (\ref{eq:Kane-Mele, real space})
is a generalization of the Kane-Mele quantum spin Hall model~\cite{Kane2005} to the case of inversion
symmetry breaking. The Hamiltonian becomes decoupled in the $\bar{\vec k}$ space of the superlattice,
$
H = \sum_{\bar{\vec{k}}} d^\dagger(\bar{\vec{k}})
h\left(\bar{\vec{k}}\right)
d(\bar{\vec{k}}),
$
where 
$
d(\bar{\vec{k}}) =  [c_{A\bar{\vec{k}}\uparrow},c_{A\bar{\vec{k}}\downarrow},c_{B\bar{\vec{k}}\uparrow},c_{B\bar{\vec{k}}\downarrow}]^T.
$
We use an overbar in $ \bar{\vec{k}} $ to indicate that the wave vector is associated with the superlattice. At the superlattice Brillouin zone corners $\bar{\vec{k}}=\tau_{z}\left(\frac{4\pi}{3\bar{a}},0\right)$,
with $\bar{a}$ the superlattice constant and $\tau_{z}=\pm1$, 
\begin{equation}
	h\left(\bar{\vec{k}}\right)=\Delta\sigma_{z}+\lambda\sigma_{z}\tau_{z}s_{z}+\xi\tau_{z}s_{z}\label{eq:Kane-Mele at valleys} \text{,}
\end{equation}
in which $\sigma_{z}=+1$ for the $ A $ sublattice or $-1$ for the $ B $ sublattice.
The pseudospin  
is a good quantum number because $ [s_{z},H]=0 $. By analogy to the spin Chern number characterizing a quantum spin Hall system~\cite{Kane2005}, a pseudospin Chern number can be defined for the optical superlattice as 
\begin{equation}
	C_{s}=\frac{1}{2}(C_{\uparrow}-C_{\downarrow}) \text{,}
\end{equation}
in which $C_{\uparrow} $ and $C_{\downarrow} $ are the Chern numbers of the $ \uparrow $ and $ \downarrow $ pseudospin branches of the eigenstates, respectively.
It is evident from \Eq (\ref{eq:Kane-Mele at valleys})  that the band gaps close at $\bar{K}$ or $\bar{K}'$ of the superlattice Brillouin
zone when $\left|\Delta \right|=\left|\lambda\right|$. For $\left|\Delta\right|>\left|\lambda\right|$, the pseudospin Chern
number is 0, and for $\left|\lambda\right|>\left|\Delta\right|$ the pseudospin
Chern number is $\pm1$. This means that the presence of the ESOC
term in the effective-mass equation (\ref{eq:corrected effective mass equation})
is able to drive a topological phase transition of the optical superlattice.

A concrete example of the ESOC-induced topological phase is now provided by numerically solving the above-mentioned superlattice, whose effective-mass theory can be furnished by \Eq (\ref{eq:Kane-Mele, real space}). The large-scale potential $ U(x,y) $ which plays the role of the ``impurity'' potential in the higher-order effective-mass equation (\ref{eq:corrected effective mass equation}) is chosen as
\begin{widetext}
\begin{equation}\label{eq: large-scale potential}
	U\left(x,y\right)=2V_{I}\left\{ \cos\left[2\pi\left(\frac{x}{A}-\frac{1}{\sqrt{3}}\frac{y}{A}\right)\right]+\cos\left[2\pi\left(\frac{x}{A}+\frac{1}{\sqrt{3}}\frac{y}{A}\right)\right]+\cos\left(\frac{4\pi}{\sqrt{3}}\frac{y}{A}\right)\right\}.
\end{equation}
\end{widetext}
This is a honeycomb lattice potential with lattice constant $ A $. Three additional laser beams can be introduced to create the large-scale potential (\ref{eq: large-scale potential}), as schematically shown by the three green arrows in \Fig \ref{fig: optical_lattice}. They are coming in with angles of incidence $ \theta $ with respect to the plane of the host lattice. These three laser beams can also be shunted from the three laser sources which generate the host lattice, using acousto-optic modulators and mirrors.  Then the relation between the lattice constants $ A $ and $ a $ is
\begin{equation}
	A = a / \sin\theta.	
\end{equation}
In numerical calculations, we use $ a $ as the length unit, bare mass $ m $ of the atom as the mass unit, and $ \hbar^2/ma^2 $ as the energy unit, i.e., we set $ \hbar = m = a = 1 $. Other parameters are chosen as $ V_{A} = 20.0 $, $ V_{B} = 20.2 $, $ V_{I} = 0.25 $, and $ A = 12 $. The lattice constant $ A = 12 $ can be realized by choosing the incident angle $ \theta = 4.78^{\circ} $. The superlattice is schematically shown in \Fig \ref{fig: superlattice_bands_sin}(a). The Hamiltonian is expanded in a plane-wave basis set. The cutoff
of the wave vectors is $38\left|\bar{\vec{b}}_{1}\right|$ and $38\left|\bar{\vec{b}}_{2}\right|$
for the two dimensions, respectively, with $\bar{\vec{b}}_{1}$and $\bar{\vec{b}}_{2}$
denoting the primitive vectors of the reciprocal lattice of the superlattice.

\begin{figure}
	\begin{centering}
		\includegraphics[width=8.5cm]{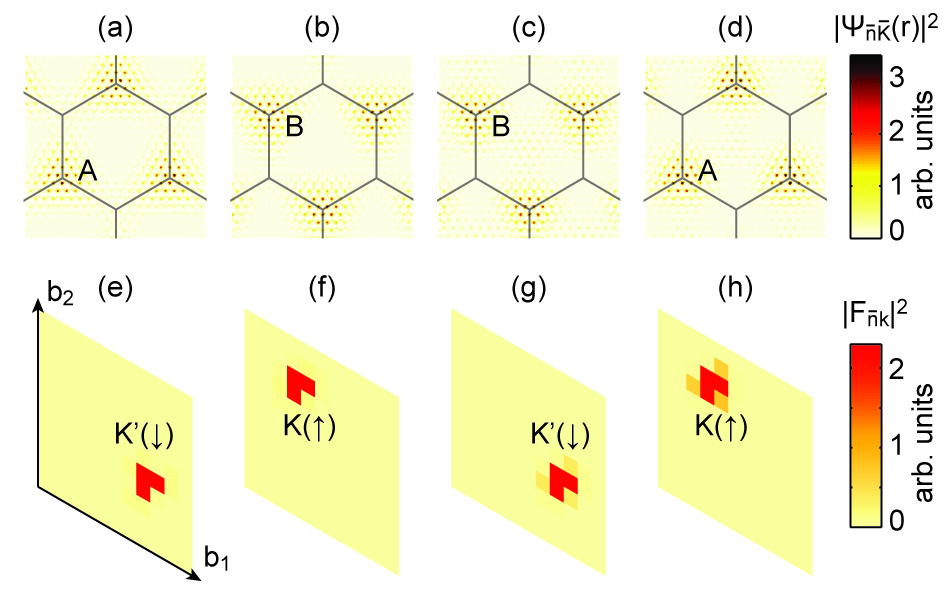}
		\par\end{centering}
	
	\protect\caption{\label{fig: Psi+Fnk_sin}(a)--(d): Spatial distributions of
		the particle densities $\left|\Psi_{\bar{n}\bar{K}}\left(\vec{r}\right)\right|^{2}$
		of the four miniband states at the $\bar{K}$ point of the superlattice Brillouin
		zone, in ascending order of energy. The black lines indicate the honeycomb superlattice. (e)--(h): Distributions of $\left|F_{\bar{n}\vec{k}}\right|^{2}=\left|\left\langle \psi_{c\vec{k}}|\Psi_{\bar{n}\bar{K}}\right\rangle\right|^{2}$
		of the four miniband states at the
		$\bar{K}$ point, in ascending order
		of energy. The parallelograms are primitive cells of the reciprocal of the host lattice.}
\end{figure}

As expected, four minibands are found in the band gap of the host lattice as shown in Fig. \ref{fig: superlattice_bands_sin}(b), originating
from the two $ 1s $ states of the $ A $ and $ B $ sublattice: $|A\uparrow\rangle$,
$|A\downarrow\rangle$, $|B\uparrow\rangle$, and $|B\downarrow\rangle$.  
To determine the topological nature of the minibands, let us first
investigate the components of the four states at the $\bar{K}$ point of the
superlattice Brillouin zone, which reflects the relative magnitudes of $ |\Delta| $ and $ |\lambda| $ in \Eq (\ref{eq:Kane-Mele at valleys}). The spatial distributions of the particle
densities $\left|\Psi_{\bar{n}\bar{K}}\left(\vec{r}\right)\right|^{2}$ of the four miniband states $ \bar{n}=1\text{--}4 $ are
calculated and shown in Figs. \ref{fig: Psi+Fnk_sin}(a)--\ref{fig: Psi+Fnk_sin}(d).
It can be seen that the particle density is localized near the trapping sites of $ U(\vec{r}) $, confirming that the minibands are indeed derived from bound states of the trapping sites. The distributions of the projection squared of
the miniband states at $ \bar{K} $ onto the Bloch states of the host lattice $\left|F_{\bar n \vec{k}}\right|^{2}=\left|\left\langle \psi_{c\vec{k}}|\Psi_{\bar{n}\bar{K}}\right\rangle\right|^{2}$, $ \bar{n}=1\text{--}4 $
are shown in Figs. \ref{fig: Psi+Fnk_sin}(e)--\ref{fig: Psi+Fnk_sin}(h). We have checked
that these four states are constructed almost entirely from the
Bloch states $ | \psi_{c\vec{k}} \rangle $ of the conduction band of the host lattice. The distributions of $\left|F_{\bar n \vec{k}}\right|^{2}$
are mostly concentrated near the two valleys at $K$ and $K'$ of the
host lattice Brillouin zone, which validates the assumptions on which the effective-mass theory is based. 

From Fig. \ref{fig: Psi+Fnk_sin} the components of the four impurity
states at $ \bar{K} $ can be readily seen, and it is found that $|A\uparrow\rangle>|B\downarrow\rangle>|B\uparrow\rangle>|A\downarrow\rangle$
by their energy. Then the order of the magnitudes of parameters in the Hamiltonian
(\ref{eq:Kane-Mele at valleys}) can be determined, namely, $-\xi<\Delta<\xi<\lambda$.
This set of parameters indicates that the impurity superlattice lies
in a topologically nontrivial regime. Calculations of the TB band structures can be found in the SM~\cite{SM}. The TB parameters lying in the topologically nontrivial regime reasonably reproduce the minibands, which indicates that the generalized Kane-Mele model (\ref{eq:Kane-Mele, real space}) is indeed reliable for the description of the minibands.

\begin{figure}
	\begin{center}
		\includegraphics[width=8.5cm]{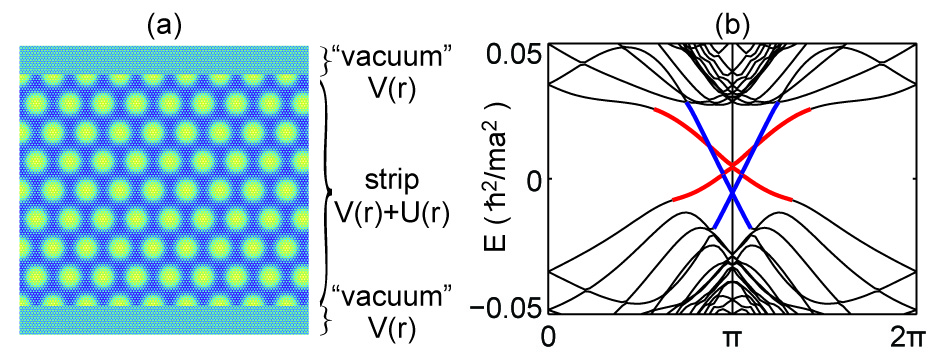}
	\end{center}
	\caption{\label{fig: edge_state_first_principles}(a) The strip model for calculation of edge states. (b) Band structure of the strip model. The gapless edge states are highlighted by red and blue lines.}
\end{figure}

The topological nature of the optical superlattice indicates the existence of topologically protected edge states, which we now confirm. We build a strip model in which the superlattice potential $ U(\vec{r}) $ is truncated to four periods of the armchair direction and put on top of the host lattice potential $ V(\vec{r}) $, as shown in \Fig \ref{fig: edge_state_first_principles}(a). Plane-wave expansion is used to solve this strip model. The band structure near the energy window of the minibands is shown in \Fig \ref{fig: edge_state_first_principles}(b). The gapless edge states as marked by red and blue lines can be readily seen. We have checked that the two pairs of states at $ k=\pi $ are indeed lying on the two edges of the strip~\cite{SM}. The existence of the gapless edge states provides concrete evidence of the topological nature of the minibands, and validates the new effective-mass equation on which the generalized Kane-Mele model is built.

In the approach to creating SOC in an optical superlattice proposed above, a remarkable point is that the minibands only experience the influence of the host lattice via the effective mass, Berry curvature, and quantum metric at the valleys. All  high-energy degrees of freedom are out of the picture. This is particularly advantageous, as it is clear from the new effective-mass equation (\ref{eq:corrected effective mass equation}) that the relative position, orientation, and commensuration between the host and superlattices need not be precisely controlled to procure the geometric effects of the host lattice. Thus, the proposed approach is flexible. For example, Tarruell \textit{et al}.~\cite{Tarruell2012} demonstrated experimentally that anisotropic massive Dirac cone dispersion can be created on an inversion symmetry-broken honeycomb lattice. Although anisotropic, this provides an equally ideal experimental host lattice for realizing the geometry-driven SOC and topological phase.

Putting the current effective-mass equation in  a more general context, a richer  variety of SOC effects can be achieved. For example, spins larger than 1/2 can be simulated on a host lattice with more than two valleys, and 3D SOC can be generated by a suitably constructed 3D host lattice. The effective SOC in this work does not flip spins. To generate a spin-flipping effective SOC, one may consider constructing a host lattice whose energy valley(s) consists of two or more degenerate bands acting as spins. Then the Berry curvature in the new effective-mass equation (\ref{eq:corrected effective mass equation}) is non-Abelian~\cite{Chang2008,Xiao10}, and its nonzero off-diagonal matrix elements can flip the spins.

In summary, the geometric effects of Bloch bands have a highly nontrivial manifestation in the quantum states and spectrum, in the presence of a perturbation breaking discrete translational symmetry. In comparison with the traditional effective-mass equation, the geometric effects lead to two relativisticlike terms in the Hamiltonian, namely, the effective SOC and effective Darwin term. The new equation provides a different perspective for understanding and tailoring wave-mechanical states associated with imperfections of a host lattice, which can be any spatially periodic system including crystals, optical lattices, artificial superstructures, and so on. As an example, we demonstrate that the effective SOC can be exploited to induce nontrivial topology in an optical superlattice. This artificial SOC on cold atoms comes from the geometry of the band structure of the optical host lattice, and does not involve the internal structure of the atoms. Hence, this scheme has a broad applicability on various kinds of atoms, and avoids the heating or atom loss problem from spontaneous emission. The new effective-mass approximation can also be applied to other artificial superlattices, which is demonstrated in detail in the SM \cite{SM}. The formalism developed herein may also be profitably extended to photonic crystals, which are governed by the Maxwell equations.

\begin{acknowledgments}
We are grateful to Di Xiao and Xibo Zhang for helpful discussions. This work is supported by the Ministry of Science and Technology of the People’s Republic of China (Grants No. 2018YFA0305601, No. 2016YFA0301004, No. 2017YFA0303302, and No. 2018YFA0305603), National Natural Science Foundation of China (Grant No. 11725415), and by the Strategic Priority Research Program of Chinese Academy of Sciences, Grant No. XDB28000000. Q.N. is supported in the U.S. by DOE (DE-FG03-02ER45958, Division of Materials Science and Engineering) on the general aspect of geometric phase effect, and by NSF (EFMA-1641101) and the Welch Foundation (F-1255).
\end{acknowledgments}



\section*{Supplementary material}

# Supplemental Material for
# "Geometric effects in the effective-mass theory and topological optical superlattices"

Chao-Kai Li,[1] Qian Niu,[2] and Ji Feng[1,3,4,]*

[1]*International Center for Quantum Materials,
School of Physics, Peking University, Beijing 100871, China*

[2]*Department of Physics, the University of Texas, Austin, TX 78712-0264, USA*

[3]*Collaborative Innovation Center of Quantum Matter, Beijing 100871, China*

[4]*CAS Center for Excellence in Topological Quantum Computation,
University of Chinese Academy of Sciences, Beijing 100190, China*

(Dated: October 8, 2018)

---

* jfeng11@pku.edu.cn

# I. DETAILS OF THE DERIVATION OF THE NEW EFFECTIVE-MASS EQUATION

We begin by the Eq. (3) of the main text

$$\varepsilon_n(\boldsymbol{k})F_{n\boldsymbol{k}} + \sum_{n'}\int_{\mathrm{BZ}}[\mathrm{d}\boldsymbol{k}]\,\langle\psi_{n\boldsymbol{k}}|U|\psi_{n'\boldsymbol{k}'}\rangle F_{n'\boldsymbol{k}'} = EF_{n\boldsymbol{k}}, \tag{1}$$

and the matrix element of the assumed slowly varying potential $U(\boldsymbol{r})$ is approximated by the Eq. (4) in the main text

$$\langle\psi_{n\boldsymbol{k}}|U|\psi_{n'\boldsymbol{k}'}\rangle \approx \frac{(2\pi)^{d/2}}{V_c}U(\boldsymbol{k}-\boldsymbol{k}')\,\langle u_{n\boldsymbol{k}}|u_{n'\boldsymbol{k}'}\rangle_{\mathrm{cell}}. \tag{2}$$

Because the expansion coefficient $F_{n\boldsymbol{k}}$ is concentrated near the band extremum position $\boldsymbol{k}_0$, only the matrix elements $\langle\psi_{n\boldsymbol{k}}|U|\psi_{n'\boldsymbol{k}'}\rangle$ with $\boldsymbol{k}\approx\boldsymbol{k}'\approx\boldsymbol{k}_0$ is relevant in Eq. (1). Thus, we can make Taylor expansion of $\langle u_{n\boldsymbol{k}}|$ and $|u_{n'\boldsymbol{k}'}\rangle$ in Eq. (2) up to second order of $\delta\boldsymbol{k} = \boldsymbol{k}-\boldsymbol{k}_0$ and $\delta\boldsymbol{k}' = \boldsymbol{k}'-\boldsymbol{k}_0$,

$$\begin{aligned}
&\langle u_{n\boldsymbol{k}}|u_{n'\boldsymbol{k}'}\rangle \\
&\approx \left(\langle u_{n\boldsymbol{k}_0}| + \langle\partial_{k_\alpha}u_{n\boldsymbol{k}}|_{\boldsymbol{k}_0}\delta k^\alpha + \frac{1}{2}\langle\partial_{k_\alpha}\partial_{k_\beta}u_{n\boldsymbol{k}}|_{\boldsymbol{k}_0}\delta k^\alpha\delta k^\beta\right) \\
&\quad\times\left(|u_{n'\boldsymbol{k}_0}\rangle + |\partial_{k_\alpha}u_{n'\boldsymbol{k}}\rangle_{\boldsymbol{k}_0}\delta k'^\alpha + \frac{1}{2}|\partial_{k_\alpha}\partial_{k_\beta}u_{n'\boldsymbol{k}}\rangle_{\boldsymbol{k}_0}\delta k'^\alpha\delta k'^\beta\right) \\
&\approx \delta_{nn'} + \langle u_{n\boldsymbol{k}}|\partial_{k_\alpha}u_{n'\boldsymbol{k}}\rangle_{\boldsymbol{k}_0}(\delta k'^\alpha - \delta k^\alpha) + \partial_{k_\alpha}\langle u_{n\boldsymbol{k}}|\partial_{k_\beta}u_{n'\boldsymbol{k}}\rangle_{\boldsymbol{k}_0}\delta k^\alpha\left(\delta k'^\beta - \delta k^\beta\right) \\
&\quad + \frac{1}{2}\langle u_{n\boldsymbol{k}}|\partial_{k_\alpha}\partial_{k_\beta}u_{n'\boldsymbol{k}}\rangle_{\boldsymbol{k}_0}\left(\delta k^\alpha\delta k^\beta + \delta k'^\alpha\delta k'^\beta - 2\delta k^\alpha\delta k'^\beta\right),
\end{aligned} \tag{3}$$

where the identity $\langle u_{n\boldsymbol{k}}|u_{n'\boldsymbol{k}}\rangle = \delta_{nn'}$ and its derivative, as well as the symmetry between the dummy indices $\alpha$ and $\beta$ are used.

In the second term of Eq. (1), the range of integration can be approximately extended from the Brillouin zone to the entire k-space, since only the small region near $\boldsymbol{k}_0$ makes a significant contribution. Inserting Eqs. (2) and (3) into the second term of Eq. (1), we have

$$\begin{aligned}
&\int_{\mathrm{BZ}}\mathrm{d}^d\boldsymbol{k}'\,\langle\psi_{n\boldsymbol{k}}|U|\psi_{n'\boldsymbol{k}'}\rangle F_{n'\boldsymbol{k}'} \\
&\approx \int_\infty \mathrm{d}^d\delta\boldsymbol{k}'\,U(\delta\boldsymbol{k}-\delta\boldsymbol{k}')\left[\delta_{nn'} + \langle u_{n\boldsymbol{k}}|\partial_{k_\alpha}u_{n'\boldsymbol{k}}\rangle_{\boldsymbol{k}_0}(\delta k'^\alpha - \delta k^\alpha)\right. \\
&\qquad\qquad\qquad\left. + \partial_{k_\alpha}\langle u_{n\boldsymbol{k}}|\partial_{k_\beta}u_{n'\boldsymbol{k}}\rangle_{\boldsymbol{k}_0}\delta k^\alpha\left(\delta k'^\beta - \delta k^\beta\right)\right.
\end{aligned}$$

$$+\frac{1}{2}\langle u_{n\bm{k}}|\partial_{k_\alpha}\partial_{k_\beta}u_{n'\bm{k}}\rangle_{\bm{k}_0}\left(\delta k^\alpha \delta k^\beta + \delta k'^\alpha \delta k'^\beta - 2\delta k^\alpha \delta k'^\beta\right)\Big]F_{n',\bm{k}_0+\delta\bm{k}'}$$

$$=\int_\infty d^d\delta\bm{k}' U(\delta\bm{k}')\Big[\delta_{nn'} + \mathrm{i}A_{nn',\alpha}(\bm{k}_0)\delta k'^\alpha + \mathrm{i}\partial_{k_\beta}A_{nn',\alpha}(\bm{k}_0)\delta k^\beta \delta k'^\alpha$$

$$-\frac{1}{2}G_{nn',\alpha\beta}(\bm{k}_0)\delta k'^\alpha \delta k'^\beta\Big]F_{n',\bm{k}_0+\delta\bm{k}-\delta\bm{k}'}, \qquad (4)$$

in which

$$A_{nn',\alpha}(\bm{k}) = \mathrm{i}\langle u_{n\bm{k}}|\partial_{k_\alpha}|u_{n'\bm{k}}\rangle \qquad (5)$$

is the Berry connection [1], and

$$G_{nn',\alpha\beta}(\bm{k}) = -\langle u_{n\bm{k}}|\partial_{k_\alpha}\partial_{k_\beta}u_{n'\bm{k}}\rangle \qquad (6)$$

With Eq. (4), we get the Schrödinger equation in k-space

$$\varepsilon_n(\bm{k}_0+\delta\bm{k})F_{n,\bm{k}_0+\delta\bm{k}} + \frac{1}{(2\pi)^{d/2}}\sum_{n'}\int_\infty d^d\delta\bm{k}' U(\delta\bm{k}')\Big[\delta_{nn'} + \mathrm{i}A_{nn',\alpha}(\bm{k}_0)\delta k'^\alpha$$

$$+\mathrm{i}\partial_{k_\beta}A_{nn',\alpha}(\bm{k}_0)\delta k^\beta \delta k'^\alpha - \frac{1}{2}G_{nn',\alpha\beta}(\bm{k}_0)\delta k'^\alpha \delta k'^\beta\Big]F_{n',\bm{k}_0+\delta\bm{k}-\delta\bm{k}'} = EF_{n',\bm{k}_0+\delta\bm{k}}. \qquad (7)$$

To transform into real space, using the Fourier transformations

$$F_{n,\bm{k}_0+\delta\bm{k}} = \frac{1}{(2\pi)^{d/2}}\int_\infty d^d\bm{r}\, e^{-\mathrm{i}\delta\bm{k}\cdot\bm{r}} F_{n\bm{k}_0}(\bm{r}) \qquad (8)$$

and

$$U(\bm{r}) = \frac{1}{(2\pi)^{d/2}}\int_\infty d^d\bm{k}\, U(\bm{k}) e^{\mathrm{i}\bm{k}\cdot\bm{r}}, \qquad (9)$$

we have

$$\varepsilon_n(\bm{k}_0+\delta\bm{k})F_{n,\bm{k}_0+\delta\bm{k}} = \frac{1}{(2\pi)^{d/2}}\int_\infty d^d\bm{r}\, e^{-\mathrm{i}\delta\bm{k}\cdot\bm{r}} \varepsilon_n(\bm{k}_0-\mathrm{i}\nabla)F_{n\bm{k}_0}(\bm{r}), \qquad (10)$$

$$\int_\infty d^d\delta\bm{k}' U(\delta\bm{k}')F_{n',\bm{k}_0+\delta\bm{k}-\delta\bm{k}'} = \int_\infty d^d\bm{r}\, e^{-\mathrm{i}\delta\bm{k}\cdot\bm{r}} U(\bm{r}) F_{n'\bm{k}_0}(\bm{r}), \qquad (11)$$

$$\int_\infty d^d\delta\bm{k}' U(\delta\bm{k}')\mathrm{i}A_{nn',\alpha}(\bm{k}_0)\delta k'^\alpha F_{n',\bm{k}_0+\delta\bm{k}-\delta\bm{k}'}$$

$$= \int_\infty d^d\bm{r}\, e^{-\mathrm{i}\delta\bm{k}\cdot\bm{r}} A_{nn',\alpha}(\bm{k}_0)\partial^\alpha U(\bm{r}) F_{n'\bm{k}_0}(\bm{r}), \qquad (12)$$

$$\int_\infty d^d\delta\bm{k}' U(\delta\bm{k}')\mathrm{i}\partial_{k_\beta}A_{nn',\alpha}(\bm{k}_0)\delta k^\beta \delta k'^\alpha F_{n',\bm{k}_0+\delta\bm{k}-\delta\bm{k}'}$$

$$= \int_\infty d^d\bm{r}\, e^{-\mathrm{i}\delta\bm{k}\cdot\bm{r}}\mathrm{i}\partial_{k_\beta}A_{nn',\alpha}(\bm{k}_0)\left[\left(-\partial^\alpha\partial^\beta U(\bm{r})\right)F_{n'\bm{k}_0}(\bm{r}) - \partial^\alpha U(\bm{r})\partial^\beta F_{n'\bm{k}_0}(\bm{r})\right], \qquad (13)$$

3

and

$$\int_\infty d^d \delta \boldsymbol{k}' U(\delta \boldsymbol{k}') \left(-\frac{1}{2}\right) G_{nn',\alpha\beta}(\boldsymbol{k}_0) \delta k'^\alpha \delta k'^\beta F_{n',\boldsymbol{k}_0+\delta\boldsymbol{k}-\delta\boldsymbol{k}'}$$
$$= \int_\infty d^d \boldsymbol{r} e^{-i\delta\boldsymbol{k}\cdot\boldsymbol{r}} \frac{1}{2} G_{nn',\alpha\beta}(\boldsymbol{k}_0) \left(\partial^\alpha \partial^\beta U(\boldsymbol{r})\right) F_{n'\boldsymbol{k}_0}(\boldsymbol{r}), \quad (14)$$

where integration by parts is used in deriving these equations. With Eqs. (10)-(14), we have

$$\sum_{n'} \Big\{ [\varepsilon_n(\boldsymbol{k}_0 - i\nabla) + U(\boldsymbol{r})]\delta_{nn'} + \partial^\alpha U(\boldsymbol{r}) \left[A_{nn',\alpha}(\boldsymbol{k}_0) + \partial_{k_\beta} A_{nn',\alpha}(\boldsymbol{k}_0)(-i\partial^\beta)\right]$$
$$+ \frac{1}{2}\left[-i\left(\partial_{k_\alpha} A_{nn',\beta}(\boldsymbol{k}_0) + \partial_{k_\beta} A_{nn',\alpha}(\boldsymbol{k}_0)\right) + G_{nn',\alpha\beta}(\boldsymbol{k}_0)\right]$$
$$\times \left(\partial^\alpha \partial^\beta U(\boldsymbol{r})\right) \Big\} F_{n'\boldsymbol{k}_0}(\boldsymbol{r}) = E F_{n\boldsymbol{k}_0}(\boldsymbol{r}) \quad (15)$$

The second line can be further simplified by

$$-i\left(\partial_{k_\alpha} A_{nn',\beta}(\boldsymbol{k}_0) + \partial_{k_\beta} A_{nn',\alpha}(\boldsymbol{k}_0)\right) + G_{nn',\alpha\beta}(\boldsymbol{k}_0)$$
$$= \partial_{k_\alpha} \langle u_{n\boldsymbol{k}} | \partial_{k_\beta} u_{n'\boldsymbol{k}} \rangle_{\boldsymbol{k}_0} + \partial_{k_\beta} \langle u_{n\boldsymbol{k}} | \partial_{k_\alpha} u_{n'\boldsymbol{k}} \rangle_{\boldsymbol{k}_0} - \langle u_{n\boldsymbol{k}} | \partial_{k_\alpha} \partial_{k_\beta} u_{n'\boldsymbol{k}} \rangle_{\boldsymbol{k}_0}$$
$$= \partial_{k_\alpha} \langle u_{n\boldsymbol{k}} | \partial_{k_\beta} u_{n'\boldsymbol{k}} \rangle_{\boldsymbol{k}_0} + \langle \partial_{k_\beta} u_{n\boldsymbol{k}} | \partial_{k_\alpha} u_{n'\boldsymbol{k}} \rangle_{\boldsymbol{k}_0}$$
$$= -\langle \partial_{k_\alpha} \partial_{k_\beta} u_{n\boldsymbol{k}} | u_{n'\boldsymbol{k}} \rangle_{\boldsymbol{k}_0} = G^\dagger_{nn',\alpha\beta}(\boldsymbol{k}_0). \quad (16)$$

In the end, we obtain the new effective-mass equation in real space:

$$\sum_{n'} \Big[ (\varepsilon_n(\boldsymbol{k}_0 - i\nabla) + U(\boldsymbol{r}))\delta_{nn'} + \nabla U(\boldsymbol{r}) \cdot \boldsymbol{A}_{nn'}(\boldsymbol{k}_0 - i\nabla)$$
$$+ \frac{1}{2} G^\dagger_{nn',\alpha\beta}(\boldsymbol{k}_0)\left(\partial^\alpha \partial^\beta U(\boldsymbol{r})\right) \Big] F_{n'\boldsymbol{k}_0}(\boldsymbol{r}) = E F_{n\boldsymbol{k}_0}(\boldsymbol{r}), \quad (17)$$

in which $A_{nn',\alpha}(\boldsymbol{k}_0 - i\nabla)$ is understood as its first order expansion

$$A_{nn',\alpha}(\boldsymbol{k}_0 - i\nabla) \approx A_{nn',\alpha}(\boldsymbol{k}_0) + \partial_{k_\beta} A_{nn',\alpha}(\boldsymbol{k}_0)(-i\partial^\beta). \quad (18)$$

## II. GAUGE TRANSFORMATION

For a one-band gauge transformation

$$|\tilde{u}_{n\boldsymbol{k}}\rangle = e^{i\theta_n(\boldsymbol{k})}|u_{n\boldsymbol{k}}\rangle, \quad (19)$$

the diagonal matrix elements $A_{n,\alpha}(\boldsymbol{k})$ and $G_{n,\alpha\beta}(\boldsymbol{k})$ transform as

$$\tilde{A}_{n,\alpha} = A_{n,\alpha} - \partial_{k_\alpha}\theta_n \quad (20)$$

and

$$\tilde{G}_{n,\alpha\beta} = G_{n,\alpha\beta} - (\partial_{k_\alpha}\theta_n)\tilde{A}_{n,\beta} - (\partial_{k_\beta}\theta_n)\tilde{A}_{n,\alpha} - (\partial_{k_\alpha}\theta_n)(\partial_{k_\beta}\theta_n) - \mathrm{i}\partial_{k_\alpha}\partial_{k_\beta}\theta_n. \quad (21)$$

Let

$$\theta_n(\boldsymbol{k}_0 + \delta\boldsymbol{k}) = \sum_\alpha A_{n,\alpha}(\boldsymbol{k}_0)\delta k^\alpha + \frac{1}{2}\sum_{\alpha\beta}\partial_{k_\alpha}A_{n,\beta}(\boldsymbol{k}_0)\delta k^\alpha \delta k^\beta, \quad (22)$$

we have

$$\tilde{A}_{n,\alpha}(\boldsymbol{k}_0 + \delta\boldsymbol{k}) = -\frac{1}{2}\sum_\beta [\partial_{k_\alpha}A_{n,\beta}(\boldsymbol{k}_0) - \partial_{k_\beta}A_{n,\alpha}(\boldsymbol{k}_0)]\delta k^\beta = \frac{1}{2}(\boldsymbol{\Omega}_n(\boldsymbol{k}_0) \times \delta\boldsymbol{k})_\alpha \quad (23)$$

and

$$\begin{aligned}\tilde{G}_{n,\alpha\beta}(\boldsymbol{k}_0) &= \frac{1}{2}\Big(\langle\partial_{k_\alpha}u_{n\boldsymbol{k}}|\partial_{k_\beta}u_{n\boldsymbol{k}}\rangle_{\boldsymbol{k}_0} - \langle\partial_{k_\alpha}u_{n\boldsymbol{k}}|u_{n\boldsymbol{k}}\rangle_{\boldsymbol{k}_0}\langle u_{n\boldsymbol{k}}|\partial_{k_\beta}u_{n\boldsymbol{k}}\rangle_{\boldsymbol{k}_0} \\ &\quad + \langle\partial_{k_\beta}u_{n\boldsymbol{k}}|\partial_{k_\alpha}u_{n\boldsymbol{k}}\rangle_{\boldsymbol{k}_0} - \langle\partial_{k_\beta}u_{n\boldsymbol{k}}|u_{n\boldsymbol{k}}\rangle_{\boldsymbol{k}_0}\langle u_{n\boldsymbol{k}}|\partial_{k_\alpha}u_{n\boldsymbol{k}}\rangle_{\boldsymbol{k}_0}\Big) \\ &= g_{n,\alpha\beta}(\boldsymbol{k}_0), \end{aligned} \quad (24)$$

where

$$\Omega_{n,\gamma}(\boldsymbol{k}_0) = \frac{1}{2}\sum_{\alpha\beta}\epsilon_{\alpha\beta\gamma}[\partial_{k_\alpha}A_{n,\beta}(\boldsymbol{k}_0) - \partial_{k_\beta}A_{n,\alpha}(\boldsymbol{k}_0)] \quad (25)$$

is the Berry curvature [1] and

$$g_{n,\alpha\beta}(\boldsymbol{k}_0) = \frac{1}{2}\left(\langle\partial_{k_\alpha}u_{n\boldsymbol{k}}|Q|\partial_{k_\beta}u_{n\boldsymbol{k}}\rangle_{\boldsymbol{k}_0} + \langle\partial_{k_\beta}u_{n\boldsymbol{k}}|Q|\partial_{k_\alpha}u_{n\boldsymbol{k}}\rangle_{\boldsymbol{k}_0}\right) \quad (26)$$

is the quantum metric [2], with $Q = 1 - |u_{n\boldsymbol{k}}\rangle\langle u_{n\boldsymbol{k}}|$. Eq. (23) holds up to order $O(\delta k)$ and Eq. (24) holds up to order $O(1)$, so that terms in Eq. (7) are up to order $O(\delta k^2)$.

With Eqs. (23) and (24), in three-dimensional space we can rewrite the one-band effective-mass equation as

$$\left[\varepsilon_n(\boldsymbol{k}_0 - i\nabla) + U(\boldsymbol{r}) - \frac{1}{2}\boldsymbol{\Omega}_n(\boldsymbol{k}_0) \cdot \nabla U \times (-\mathrm{i}\nabla) \right. \\ \left. + \frac{1}{2}g_{n,\alpha\beta}(\boldsymbol{k}_0)\left(\partial^\alpha\partial^\beta U(\boldsymbol{r})\right)\right]F_{n\boldsymbol{k}_0}(\boldsymbol{r}) = EF_{n\boldsymbol{k}_0}(\boldsymbol{r}). \quad (27)$$

This form has the advantage that both the Berry curvature and the quantum metric are gauge invariant quantities, which is favorable for numerical calculations.

## III. BAND STRUCTURE AND BERRY CURVATURE DISTRIBUTION OF THE HOST LATTICE

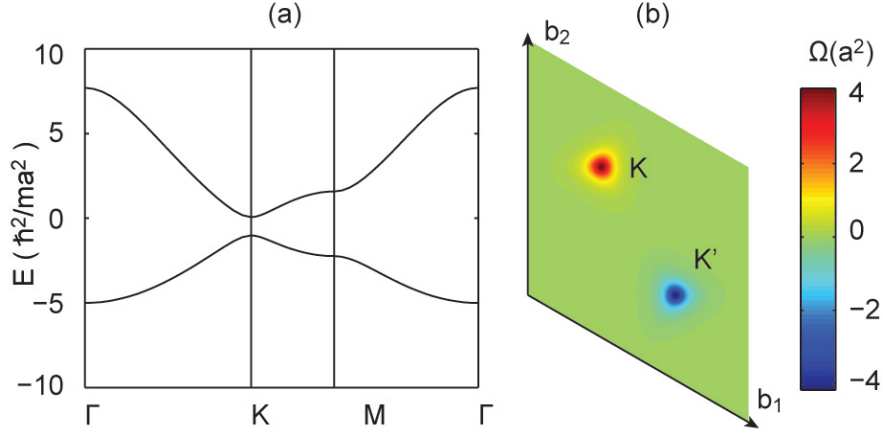

Fig. S1. (a) The lowest two energy bands of the honeycomb optical host lattice. (b) The Berry curvature distribution of the second lowest band (conduction band) of the host lattice.

## IV. CALCULATION OF $\langle \psi_{n\bm{k}} | \Psi_{\bar{n}\bar{\bm{k}}} \rangle$

In the main text, we analyzed the distribution of

$$F_{\bar{n}\bm{k}} = \langle \psi_{c\bm{k}} | \Psi_{\bar{n}\bar{K}} \rangle, \tag{28}$$

where $|\psi_{c\bm{k}}\rangle$ is the Bloch wave function of the conduction band of the host lattice, and $|\Psi_{\bar{n}\bar{K}}\rangle$ is the Bloch wave function at $\bar{K}$ point of the $\bar{n}$th band of the optical superlattice. It is a special case of the inner product $\langle \psi_{n\bm{k}} | \Psi_{\bar{n}\bar{\bm{k}}} \rangle$. We use an overbar for the quantities of the optical superlattice.

The two Bloch wave functions can be respectively expanded by plane waves as

$$\psi_{n\bm{k}}(\bm{r}) = \frac{1}{\sqrt{V}} \sum_{\bm{G}} c_{n\bm{k}}(\bm{G}) e^{i(\bm{k}+\bm{G})\cdot\bm{r}} \tag{29}$$

and

$$\Psi_{\bar{n}\bar{\bm{k}}}(\bm{r}) = \frac{1}{\sqrt{V}} \sum_{\bar{\bm{G}}} \bar{c}_{\bar{n}\bar{\bm{k}}}(\bar{\bm{G}}) e^{i(\bar{\bm{k}}+\bar{\bm{G}})\cdot\bm{r}}, \tag{30}$$

6

where $V$ is the volume of the lattice. The inner product can be calculated as

$$\langle \psi_{n\boldsymbol{k}} | \Psi_{\bar{n}\bar{\boldsymbol{k}}} \rangle = \int_{\text{crystal}} d^d \boldsymbol{r} \frac{1}{V} \sum_{\boldsymbol{G}\bar{\boldsymbol{G}}'} c_{n\boldsymbol{k}}^*(\boldsymbol{G}) \bar{c}_{\bar{n}\bar{\boldsymbol{k}}}(\bar{\boldsymbol{G}}') e^{i(\bar{\boldsymbol{k}}-\boldsymbol{k}+\bar{\boldsymbol{G}}'-\boldsymbol{G})\cdot\boldsymbol{r}}$$
$$= \sum_{\boldsymbol{G}\bar{\boldsymbol{G}}'} c_{n\boldsymbol{k}}^*(\boldsymbol{G}) \bar{c}_{\bar{n}\bar{\boldsymbol{k}}}(\bar{\boldsymbol{G}}') \delta_{\bar{\boldsymbol{k}}-\boldsymbol{k}+\bar{\boldsymbol{G}}'-\boldsymbol{G},0}. \tag{31}$$

Therefore, the inner product vanishes unless $\boldsymbol{k} - \bar{\boldsymbol{k}} = \bar{\boldsymbol{G}}$, where $\bar{\boldsymbol{G}}$ is a reciprocal lattice vector of the optical superlattice. In this case,

$$\langle \psi_{n\boldsymbol{k}} | \Psi_{\bar{n}\bar{\boldsymbol{k}}} \rangle = \delta_{\boldsymbol{k},\bar{\boldsymbol{k}}+\bar{\boldsymbol{G}}} \sum_{\boldsymbol{G}} c_{n\boldsymbol{k}}^*(\boldsymbol{G}) \bar{c}_{\bar{n}\bar{\boldsymbol{k}}}(\boldsymbol{G}+\bar{\boldsymbol{G}}). \tag{32}$$

## V. SPATIAL DISTRIBUTION OF EDGE STATES

Fig. S2 shows the spatial distributions of the edge states of the strip at $k = \pi$ (see Fig. 4(b) in the main text). It is seen that the two pairs of edge states marked by blue and red lines in Fig. 4(b) of the main text are indeed localized at the opposite edges of the strip.

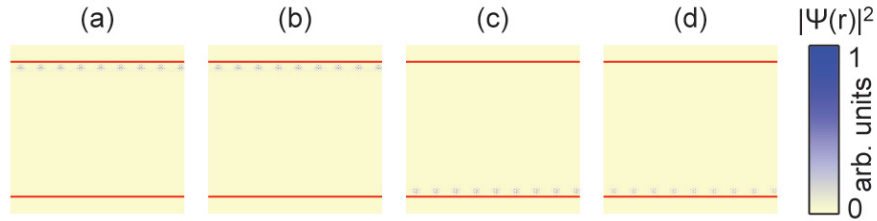

Fig. S2. (a)-(d) show the spatial distributions of the four edge states at $k = \pi$ respectively, with energy from low to high. The two red lines mark the position of the strip edge. The potential energy of the strip model is shown in Fig. 4(a) in the main text.

## VI. TIGHT-BINDING BAND STRUCTURES

We find that the set of parameters $t = 0.044$, $\Delta = -0.004975$, $\lambda = 0.03273$ and $\xi = 0.01472$ reasonably reproduce the minibands, as shown in Fig. S3(a). This set of parameters lies in the topologically non-trivial regime of the generalized Kane-Mele Hamiltonian. For this TB model we also build a 100-site-wide ribbon along the zigzag direction. The energy bands of this ribbon are shown in Fig. S3(b). There are two pairs of gapless bands inside

7

the bulk gap corresponding to edge states on the two sides of the ribbon, which are non-degenerate because the system has no inversion symmetry. The gapless edge states attest to the topological nature of the model system, confirming that the geometric effect induced ESOC can indeed be exploited to create a topologically non-trivial phase.

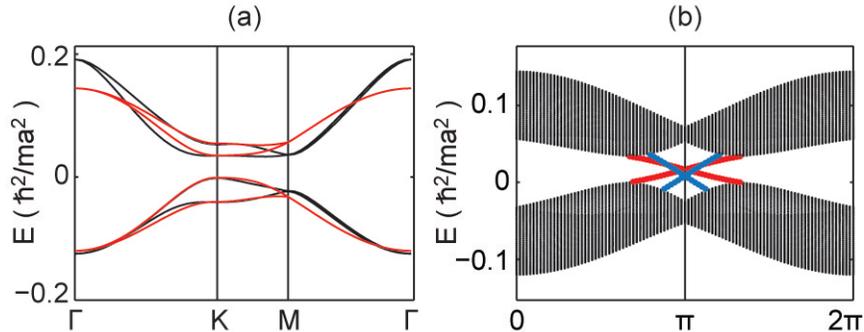

Fig. S3. (a) The four calculated minibands as in Fig. 2(b) of the main text (black) and the bands of generalized Kane-Mele TB model (red). (b) Band structure of the strip. The red and the blue lines are two pairs of gapless edge states located on the two sides of the strip.

## VII.  $\mathbb{Z}_2$ TOPOLOGICAL INVARIANT

In the discussions in the main text, the pseudospin is treated as a good quantum number, which allows us to define a pseudospin Chern number. When intervalley scattering is also considered, the pseudospin is no longer a good quantum number. However, we can define an anti-unitary operator analogous to the time-reversal operator

$$\Theta = -\mathrm{i}s_y K \tag{33}$$

in which $s_y$ is the y Pauli matrix for pseuodospin, and $K$ is the complex conjugation operator. One can verify that $\Theta$ commutes with the effective-mass Hamiltonian (5) in the main text. So long as the intervalley coupling also commutes with $\Theta$, then under a time-reversal constraint of the gauge, a $\mathbb{Z}_2$ topological invariant can be defined [3]

$$\mathbb{Z}_2 = \frac{1}{2\pi}\left[\oint_{\partial\tau_{1/2}} d\ell\ \bar{\boldsymbol{A}} - \int_{\tau_{1/2}} d\tau\ \bar{\Omega}\right] \mod 2 \tag{34}$$

8

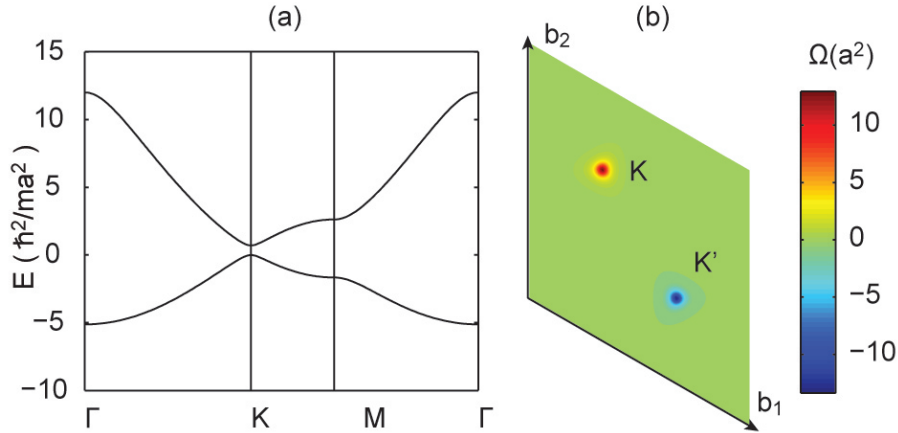

Fig. S4. (a) The lowest two energy bands of the honeycomb host lattice made up of muffin-tin potential wells. (b) The Berry curvature distribution of the second lowest band (conduction band) of the host lattice.

where $\bar{\boldsymbol{A}}$ and $\bar{\boldsymbol{\Omega}}$ are the traces of the $U(N)$ Berry connection and Berry curvature matrix calculated from the envelope function $F_{\bar{n}\bar{\boldsymbol{k}}}(\boldsymbol{r})$.

## VIII. NUMERICAL SIMULATIONS OF MUFFIN-TIN POTENTIAL

In this section, we will explore the possibility of realizing the novel 2D topological superlattice introduced in the main text in the field of materials science. The role of host lattice can be played by monolayer valleytronics materials such as BN or MoS$_2$ [4, 5]. The superlattice potential can be achieved by fabricating a periodic array of holes or bumps on the substrate, on which the monolayer material as the host lattice is subsequently transferred. This procedure is well in the reach of standard nanofabrication. For example, in Ref. [6] the authors applied this procedure to graphene in order to test its mechanical properties, and in Ref. [7] the monolayer MoS2 was placed on SiO2 substrate with periodic array of holes. The experimental setup is shown schematically in Fig. S5(a).

The host lattice can be modeled by honeycomb muffin-tin potential, which consists of finite circular wells on each lattice site. The diameter of the muffin-tin is $a/\sqrt{3}$, $a$ being the lattice constant. Here and after, $\hbar^2/ma^2$ will be used as energy units, in which $m$ is the mass of electron. The depth of potential well on the A sublattice is $u_A = -50$, and that on

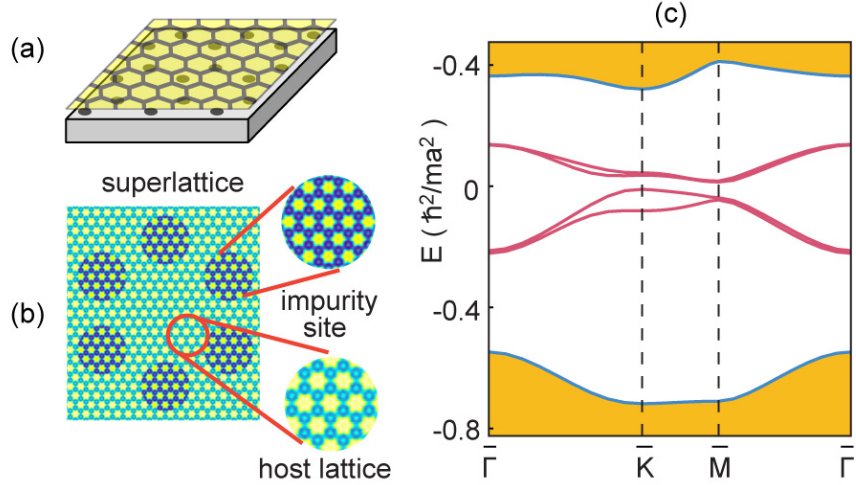

Fig. S5. (a) Schematic of an experimental setup. A sheet of monolayer material (the yellow sheet) is placed on a substrate patterned with periodic array of holes or bumps (black dots). (b) Schematic of the impurity superlattice potential. (c) Four impurity bands (red) are formed in the band gap of the host lattice.

the B sublattice is $u_B = -51$. Fig. S4 shows the band structure of the host lattice and the Berry curvature distribution of the conduction band (the upper band). They resemble the results of the optical lattice.

The holes or bumps on the substrate can also be arranged into a honeycomb lattice, referred to as the impurity lattice, and also modeled by muffin-tin potential. The lattice constant of the impurity lattice is chosen as 12 times that of the host lattice. The diameter of the muffin-tin is 8 times that of the host lattice, and has a relatively shallow depth $u_I = -2$. The total superlattice potential is schematically shown in Fig. S5(b). In solving the band structure of the superlattice, the Hamiltonian is expanded in a plane wave basis set. The cutoff of the wave vectors is $38\left|\bar{\boldsymbol{b}}_1\right|$ and $38\left|\bar{\boldsymbol{b}}_2\right|$ for the two dimensions, respectively, with $\bar{\boldsymbol{b}}_1$ and $\bar{\boldsymbol{b}}_2$ denoting the primitive vectors of the reciprocal lattice of the superlattice.

Figs. S5-S7 are the results of the calculations similar to those of the optical superlattice. The tight-binding parameters are $t = 0.0583$, $\Delta = -0.015$, $\lambda = 0.045$, $\xi = 0.02$. The analyses of these results are the same as in the optical lattice example in the main text. The energetic ordering of the superlattice pseudospin in Fig. S6 and the gapless edge states in Fig. S7(b) show clearly that the superlattice is topologically non-trivial. This demonstrates

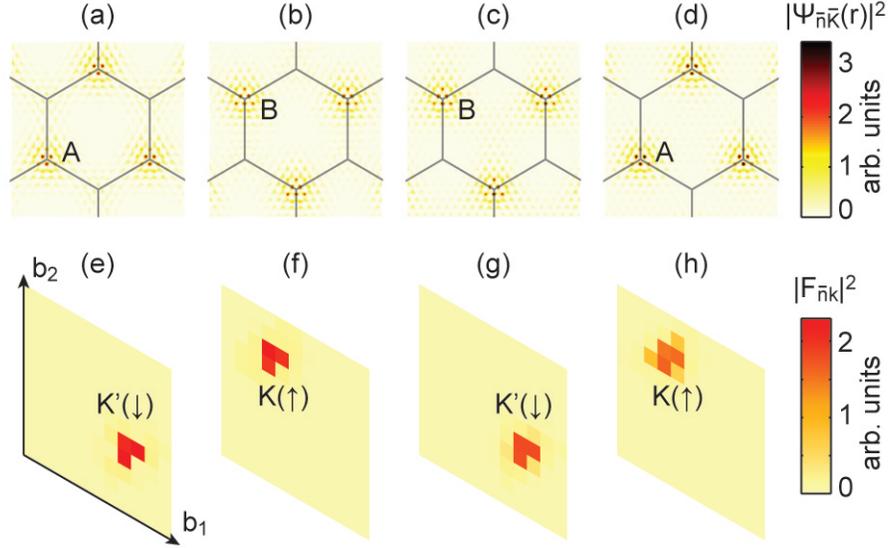

Fig. S6. (a)-(d): Spatial distributions of the particle densities $|\Psi_{\bar{n}\bar{K}}(r)|^2$ of the four impurity states at $\bar{K}$ point of the superlattice Brillouin zone, in ascending order of energy. The gray lines indicate the honeycomb superlattice. (e)-(h): Distributions of $|F_{\bar{n}k}|^2 = |\langle \psi_{ck}|\Psi_{\bar{n}\bar{K}}\rangle|^2$ of the four impurity states at $\bar{K}$ point, in ascending order of energy. The parallelograms are primitive cells of the reciprocal of the host lattice. From (a)-(h), the order of $|A \uparrow\rangle > |B \downarrow\rangle > |B \uparrow\rangle > |A \downarrow\rangle$ can be derived.

the attractive possibility of exploiting the effective spin-orbit coupling derived in this work in artificial superlattice on two-dimensional materials.

## IX. REMARKS ON THE ABSENCE OF GEOMETRIC EFFECTS IN TRADITIONAL EFFECTIVE-MASS EQUATION

We note that Ref. [8] uses an approach different from ours in deriving the effective-mass equation. we use the Bloch states $|\psi_{n\boldsymbol{k}}\rangle = e^{i\boldsymbol{k}\cdot\boldsymbol{r}}|u_{n\boldsymbol{k}}\rangle$ as the orthonormal and complete basis to expand the wavefunctions in the presence of impurity. In Ref. [8], another orthonormal basis is used, namely

$$|\chi_{n\boldsymbol{k}}\rangle = e^{i\boldsymbol{k}\cdot\boldsymbol{r}}|u_{n\boldsymbol{k}_0}\rangle, \quad (35)$$

where $\boldsymbol{k}_0$ is the location of the energy valley.

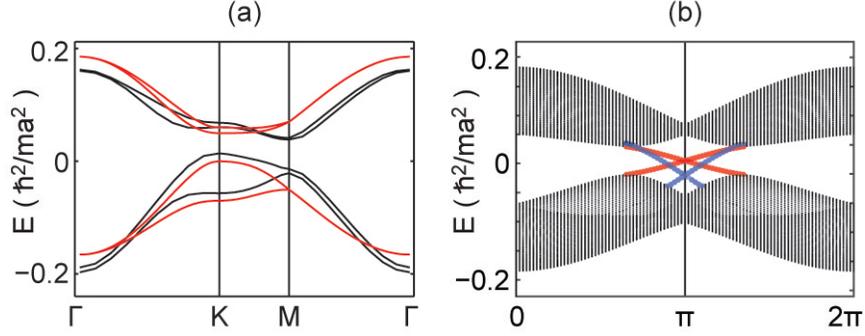

Fig. S7. (a) The four calculated impurity bands as in Fig. S5(c) (black) and the bands of generalized Kane-Mele tight-binding model (red). (b) Edge states of the strip. The red and the blue lines are two pairs of gapless edge states located on the two sides of the strip.

The two basis can be related by unitary transformations,

$$\langle\psi_{n\bm{k}}| = \sum_{n_1\bm{k}_1}(\mathrm{e}^{-S})_{n\bm{k},n_1\bm{k}_1}\langle\chi_{n_1\bm{k}_1}|, \tag{36}$$

$$|\psi_{n'\bm{k}'}\rangle = \sum_{n_2\bm{k}_2}|\chi_{n_2\bm{k}_2}\rangle(\mathrm{e}^{S})_{n_2\bm{k}_2,n'\bm{k}'}, \tag{37}$$

in which $S$ is a matrix whose elements are given by Eq. (II.31) of Ref. [8]. Then,

$$\langle\psi_{n\bm{k}}|U|\psi_{n'\bm{k}'}\rangle = \sum_{n_1\bm{k}_1 n_2\bm{k}_2}(\mathrm{e}^{-S})_{n\bm{k},n_1\bm{k}_1}\langle\chi_{n_1\bm{k}_1}|U|\chi_{n_2\bm{k}_2}\rangle(\mathrm{e}^{S})_{n_2\bm{k}_2,n'\bm{k}'} = (\mathrm{e}^{-S}\underline{\underline{U}}\mathrm{e}^{S})_{n\bm{k},n'\bm{k}'}, \tag{38}$$

where $\underline{\underline{U}}$ is the matrix whose matrix elements are $\langle\chi_{n_1\bm{k}_1}|U|\chi_{n_2\bm{k}_2}\rangle$. Expanding to the second order of $S$, we obtain

$$\langle\psi_{n\bm{k}}|U|\psi_{n'\bm{k}'}\rangle \approx \left(\underline{\underline{U}} - [S,\underline{\underline{U}}] + \frac{1}{2}[S,[S,\underline{\underline{U}}]]\right)_{n\bm{k},n'\bm{k}'}. \tag{39}$$

According to the arguments in the paragraph following Eq. (II.33) in Ref. [8], both $[S,\underline{\underline{U}}]$ and $[S,[S,\underline{\underline{U}}]]$ are neglected. Then with Eq. (II.21) of Ref. [8],

$$\langle\psi_{n\bm{k}}|U|\psi_{n'\bm{k}'}\rangle \approx (\underline{\underline{U}})_{n\bm{k},n'\bm{k}'} = \delta_{nn'}U(\bm{k}-\bm{k}'). \tag{40}$$

This is equivalent to taking

$$\langle u_{n\bm{k}}|u_{n'\bm{k}'}\rangle \approx \delta_{nn'} \tag{41}$$

12

in Eq. (4) of the main text (up to a constant factor due to different conventions of Fourier transformation), which results in the absence of geometric effects as is mentioned in the main text.



\end{document}